\documentclass{ws-rv9x6}
\usepackage{subfigure} % required only when side-by-side/subfigures are used
\usepackage{ws-rv-van} % numbered citation/references (default)
\usepackage{graphicx}
\usepackage{epsfig}
\usepackage{psfig}
\makeindex
\begin{document}

\chapter[Extreme Events in Nonlinear Lattices]
        {Extreme Events in Nonlinear Lattices}\label{ra_ch1}

\author[G. P. Tsironis$^{1,2}$, N. Lazarides$^{1,2}$, A Maluckov$^{3}$ and Lj. Had\v zievski$^{3}$]
       {G. P. Tsironis$^{1,2}$\footnote{e-mail: gts@physics.uoc.gr}, N. Lazarides$^{1,2}$, A Maluckov$^{3}$ and Lj. Had\v zievski$^{3}$}
%\index[aindx]{Author, F.} % or \aindx{Author, F.}
%\index[aindx]{Author, S.} % or \aindx{Author, S.}

\address{
$^1$Department of Physics, University of Crete, P.O. Box 2208,
71003 Heraklion, Greece, \\
$^2$Institute of Electronic Structure and Laser,
Foundation for Research and Technology-Hellas, P.O. Box 1527,
71110 Heraklion, Greece \\
$^3$P$\star$ GROUP, Vin{\v c}a Institute of Nuclear Sciences,
University of Belgrade, P.O. Box 522, 11001 Belgrade, Serbia}

\begin{abstract}
The spatiotemporal complexity induced by perturbed initial
excitations through the development of modulational instability in
nonlinear lattices with or without disorder, may lead to the
formation of very high amplitude, localized transient structures
that can be named as extreme events. We analyze the statistics of
the appearance of these collective events in two different
universal lattice models; a one-dimensional nonlinear model that
interpolates between the intergable Ablowitz-Ladik (AL) equation
and the nonintegrable discrete nonlinear Schr\"odinger (DNLS)
equation, and a two-dimensional disordered DNLS equation. In both
cases, extreme events arise in the form of discrete rogue waves as
a result of nonlinear interaction and rapid coalescence between
mobile discrete breathers. In the former model, we find power-law
dependence of the wave amplitude distribution and significant
probability for the appearance of extreme events close to the
integrable limit. In the latter model, more importantly, we find a
transition in the the return time probability of extreme events
from exponential to power-law regime. Weak nonlinearity and
moderate levels of disorder, corresponding to weak chaos regime,
favor the appearance of extreme events in that case.
\end{abstract}
%\markright{Customized Running Head for Odd Page} % default is chapter title.
\body

%%%%%%%%%%%%%%%%%%%%%%%%%%%%%%%%%%%%%%%%%%%%%%%%%%%%%%%%%%%%%%%%%%%%%%%%%%%%%%%%
\section{Introduction}\label{ra_sec1}
The inspiring work of John S. Nicolis on hierarchical systems has shown the 
significance and role of the linkage of different scales \cite{JSNicolis1986a}.  
Rogue waves may be 
seen as a form of an extreme yet emergent property of a complex system attributed 
to multiple scale hierarchies.  
Rogue or freak waves are isolated, gigantic water waves that have been observed 
to appear suddenly in relatively calm seas and disappear without a trace
\cite{Akhmediev2009a}.
Although rare, the probability of appearance of these {\em extreme events} (EE),
loosely defined as highly intense, spatially localized and temporally transient
structures \cite{Albeverio2006}, seems to be much higher than that
expected from normal, Gaussian statistics. Their theoretical
analysis has been traditionally linked to nonlinearities and/or
randomness in the water wave equations
\cite{Pelinovsky2000,Kharif2003,Shukla2006,Ruban2007,Eliasson2010,Onorato2011}.
Recently, super rogue waves have been observed in a water-wave
tank due to nonlinear focusing of the local wave amplitude
\cite{Chabchoub2012}. The occurrence of EEs is not however limited
to water waves; in physics, in particular, EEs have been observed
in a variety of systems, ranging from optical fibers
\cite{Solli2007,Hammani2008,Aalto2010,Arecchi2011}, nonlinear
optical cavities \cite{Montina2009}, superfluid $_4$He
\cite{Ganshin2009}, and laser pulse filamentation
\cite{Kasparian2009,Majus2011}, to capillary waves
\cite{Shats2010}, space plasmas \cite{Ruderman2010}, optically
injected semiconductor laser \cite{Bonatto2011}, and mode-locked
fiber lasers operating in a strongly dissipative regime
\cite{Lecaplain2012}. The existence of EEs has also been predicted
for Bose-Einstein condensates \cite{Bludov2009a}, arrays of
optical waveguides \cite{Bludov2009b}, and soft glass photonic
crystal fibers \cite{Buccoliero2011}.

In nonlinear systems, EEs may appear because of the development of
Benjamin-Feir (modulational) instability (MI) for certain types of
nonlinearity \cite{White1998}. The theoretical investigations on
EEs follow different paths; one approach adopts the nonlinear
Schr\"odinger (NLS) equation as a universal model and emphasize
the mechanisms generating short-lived soliton-like modes
\cite{Shukla2006,Osborne2010,Eliasson2010,Onorato2011,Baronio2012,Zakharov2013}.
In this approach EEs are singular events localized in space that
may be solutions of NLS-type equations with low-order (viz. cubic)
nonlinearities. For example, excitations in the form of Akhmediev
breathers \cite{Akhmediev1986,Dysthe1999,Voronovich2008} and
Peregrine solitons \cite{Henderson1999} may be formed and
coalesce, resulting in the generation of EEs. 
Other approaches, following techniques and concepts of hydrodynamics, go beyond the
cubic nonlinearity in an attempt to retain some of its complexity
\cite{Zakharov2010,Zakharov2012,Bandelow2012} and focus primarily on waves 
in continuous media; however, a wide class of interesting problems involves
wave propagation in discrete periodic media forming lattices \cite{Hennig1999}. 
Notably, MI may also develop in nonlinear lattices
\cite{Kivshar1994}, and then discrete NLS (DNLS) models are
relevant to the wave propagation in such systems as, e.g., in
nonlinear waveguide arrays. It should be noted here that
nonlinearity is not a necessary condition for the appearance of
EEs; experiments in optics and microwaves indicate that EEs may be
triggered in linear systems due to some kind of randomness
\cite{Hohman2010,Arecchi2011}. A recent review on rogue waves and
their generating mechanisms in different physical contexts is
given in Ref. \cite{Onorato2013}. Substantial research efforts
have been devoted the last few years to clarify issues related to
the probability distributions of EEs, the effect of initial
conditions, and the role of nonlinearity and/or disorder. The
issue of the interplay between disorder and nonlinearity, often
simultaneously encountered in nature and laboratory experiments,
and how it affects the probability distribution of EEs, is of
particular importance.

Nonlinear lattices form a unique workplace where several processes
of different physical nature appear simultaneously and affect
their dynamical properties \cite{Flach2008}. They constitute
prototypical systems of high spatiotemporal complexity that can be
investigated theoretically and experimentally, providing a wealth
of information on the physics of extended complex systems. The
presence of quenched disorder, introduces additionally a mechanism
for local symmetry breaking that affects their long-term dynamics
\cite{Molina1994,Kopidakis2000,Kopidakis2008,Flach2009,Pikovsky2008}.
Self-organization is one particular feature of nonlinear lattices
that is connected to the possibility for the appearance of EEs. In
this aspect, the MI plays an important role as an "intrinsic
noise" in triggering self-organization, as has been indicated by
John S. Nicolis and coworkers a long time ago
\cite{JSNicolis1975}. In this chapter we review some aspects of
the role of integrability and the interplay between disorder and
nonlinearity in one- and two-dimensional lattices described by
discrete nonlinear equations. In the next section, using a
one-dimensional discrete nonlinear model that interpolates between
the DNLS equation and the 
Ablowitz-Ladik (AL) equation by varying a single control
parameter, we obtain the optimal regime for EE generation and
their probability distribution \cite{Maluckov2009}. While the
production of EEs in nonlinear systems is mediated by MI, the
subsequent evolution reveals complex behavior and their
probability of appearance depends on the interplay of nonlinearity
and/or disorder, as well as the degree of integrability of the
system. We found that integrability properties of the lattice do
play a role in the probability of appearance of EEs
\cite{Maluckov2009}, and that the optimal regime for EE appearance
is close to the integrable limit. Importantly, the wave height
amplitude distributions match to power-law functions. A broader
perspective is obtained in section 3 through a physically
realizable model, viz. that of the two-dimensional DNLS equation
in the presence of disorder of the Anderson type
\cite{Schwartz2007,Lahini2008}. In that case, the optimal regime
for EE appearance requires weak nonlinearity and moderate levels
of disorder \cite{Maluckov2013}. Furthermore, we find a transition
in the the return time probability of EEs from exponential to
power-law regime, related to a corresponding transition of the
system from strong to weak chaos. Thus, the investigation of  both
models consistently leads to the more general conclusion that the
enhancement of probability of appearance of EEs is related to weak
chaos, since nearly integrable, modulationally unstable systems
may easily fall into a weakly chaotic state.

\section{Integrability versus non-integrability}
We consider the following model, often referred to as the Salerno
model \cite{Salerno2001}
\begin{eqnarray}
\label{eq1}
   i\frac{d\psi_n}{dt}=
   -(1+\mu |\psi_n|^2)(\psi_{n+1}+\psi_{n-1})
  -\gamma |\psi_n|^2\psi_{n}
\end{eqnarray}
where  $\mu$ and $\gamma$ are two nonlinearity parameters and $n=1,2,3,...,N$.
When
$\mu =0$ the model becomes the non-integrable DNLS equation while for $\gamma =0$
it reduces to the integrable AL equation. The norm $P_N$ and the Hamiltonian $H$ of
the model Eq. (\ref{eq1}), given respectively by
\begin{eqnarray}
\label{eq2}
   P_N=\frac{1}{\mu}\sum_n \ln{|1+\mu|\psi_n|^2|},
\end{eqnarray}
and
\begin{eqnarray}
\label{eq3}
   H=\sum_n \left[\frac{\gamma}{\mu^2} \ln|1+\mu|\psi_n|^2|
    -\frac{\gamma}{\mu} |\psi_n|^2 -2 Re[\psi_n\psi_{n+1}^*]  \right] ,
\end{eqnarray}
are both conserved quantities. Eq. (\ref{eq1})
exhibits MI that may lead to stationary solutions in
the form of discrete breathers (DBs)\cite{Flach2008}, i.e., periodic and spatially
localized nonlinear excitations. The DBs thus generated appear in random
positions in the lattice and they interact with each other,
with the high-amplitude DBs absorbing the low-amplitude ones.
After sufficient time, only a small number of high-amplitude DBs that are
pinned at particular lattice sites are left, which form virtual bottlenecks
that slow down the relaxation processes in the lattice
\cite{Tsironis1996,Rasmussen2000}.
Transient DBs, however, provide an attractive model of EEs in lattices
\cite{Dysthe1999}.
The possibility of MI development in Eq. (\ref{eq1}) can be investigated within
linear stability analysis of its the plane wave solutions modulated by small
phase and amplitude perturbations \cite{Maluckov2007}.
The relative strength of the on-site and the nonlinear interaction terms,
which is affected by the parameters $\mu$ and $\gamma$,
may change MI properties and, consequently, the conditions for transient
localization. For later convenience, the variables $\psi_n$ in Eq. (\ref{eq1})
are rescaled according to $\psi_n=\xi_n/\sqrt{\mu}$, that results in the
equation
\begin{eqnarray}
\label{eq4}
  i\frac{d\xi_n(t)}{dt}=
    -(1 +|\xi_n(t)|^2)(\xi_{n+1}+\xi_{n-1}) -\Gamma |\xi_n(t)|^2\xi_{n} ,
\end{eqnarray}
where $\Gamma=\gamma/\mu$. Therefore, the whole two-dimensional parameter space
$(\gamma,\mu)$ can be scaled by $\mu=1$, that leaves $\gamma$ as a free parameter.
With that scaling, the DNLS limit is reached for very large values of $\Gamma$.
However, the exact DNLS limit $\mu=0$ has to be calculated separately.
%%%%%%%%%%%%%%%%%%%%%%figure1%%%%%%%%%%%%%%%%%%%%%%%%%%%%%%%%%%%%%%%%%%%%%%%%%%%
\begin{figure}[h]
\includegraphics[angle=0, width=1.0\linewidth]{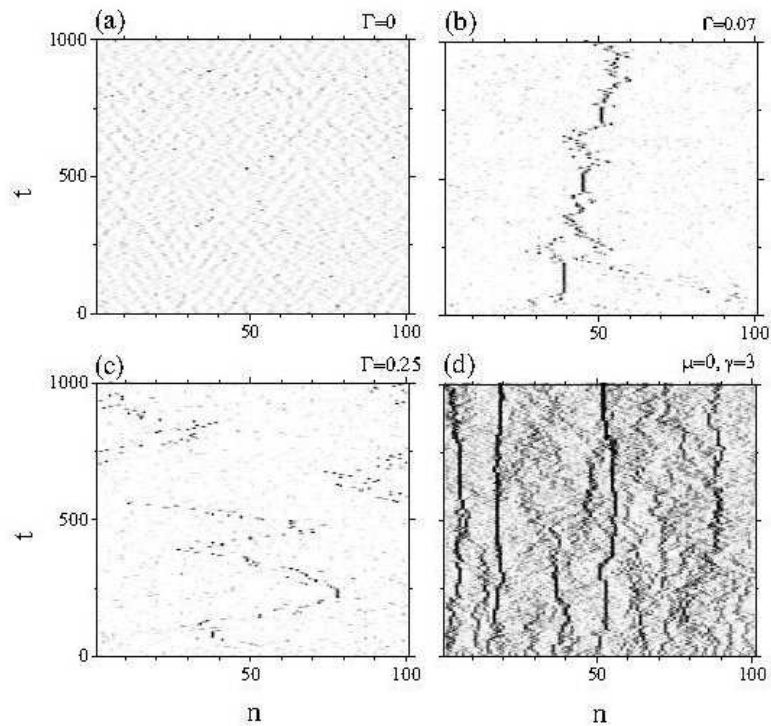}
\caption{\label{Fig. 1} Evolution of the scaled amplitudes
$|\xi_n|$ for a lattice of size $N=101$, with $\Gamma$ ($\mu$ and
$\gamma$ in the DNLS case), is shown on the figure. The initial
conditions for all cases are $\xi_n=1$ for any $n$ (uniform) plus
a small amount of white noise. }
\end{figure}
%%%%%%%%%%%%%%%%%%end-figure1%%%%%%%%%%%%%%%%%%%%%%%%%%%%%%%%%%%%%%%%%%%%%%%%%%%

Eqs. (\ref{eq4}) are integrated with a $6-$th order Runge-Kutta algorithm with
fixed time-stepping $\Delta t=10^{-4}$. Periodic boundary conditions are used
throughout this section, while the system is initiated with a uniform function
$\xi_n =1$ for any $n$, which is linearly unstable. A small amount of white
noise was added to the initial condition to accelerate the development of MI.
Different choices of initial conditions give similar results.
Variation of the parameter $\Gamma$ reveals three different regimes illustrated
in the spatiotemporal patterns shown in Fig. \ref{Fig. 1}.

(i) For the completely integrable AL lattice ($\Gamma=0$), DBs are mobile and
essentially noninteracting; as a result, the formation of EEs in this
regime is insignificant (Fig. \ref{Fig. 1}a).

(ii) In the vicinity of the AL limit, i.e. for  small values of
$\Gamma$ ($\sim 0.1$), the onset of weak interaction between
localized modes that can be observed leads to a significant
increase of EE formation (Figs. \ref{Fig. 1}b,c). In this regime
the mobility of DB excitations is rather high, indicating that DB
merging could be responsible for the appearance of EEs.

(iii) For $\Gamma >> 0.1$, the dominant behavior is of the DNLS-type,
exhibiting localized structures generated through MI that are subsequently
pinned at particular lattice sites (Fig. \ref{Fig. 1}d).
%%%%%%%%%%%%%%%%%%%%%%figure2-3%%%%%%%%%%%%%%%%%%%%%%%%%%%%%%%%%%%%%%%%%%%%%%%%%
\begin{figure}[h!]
\centerline{
  \minifigure[
The height probability density $P_h (h)$ for several values of
$\Gamma$. The dotted curve corresponds to the DNLS limit (with $\gamma=6$).
]
     {\epsfig{figure=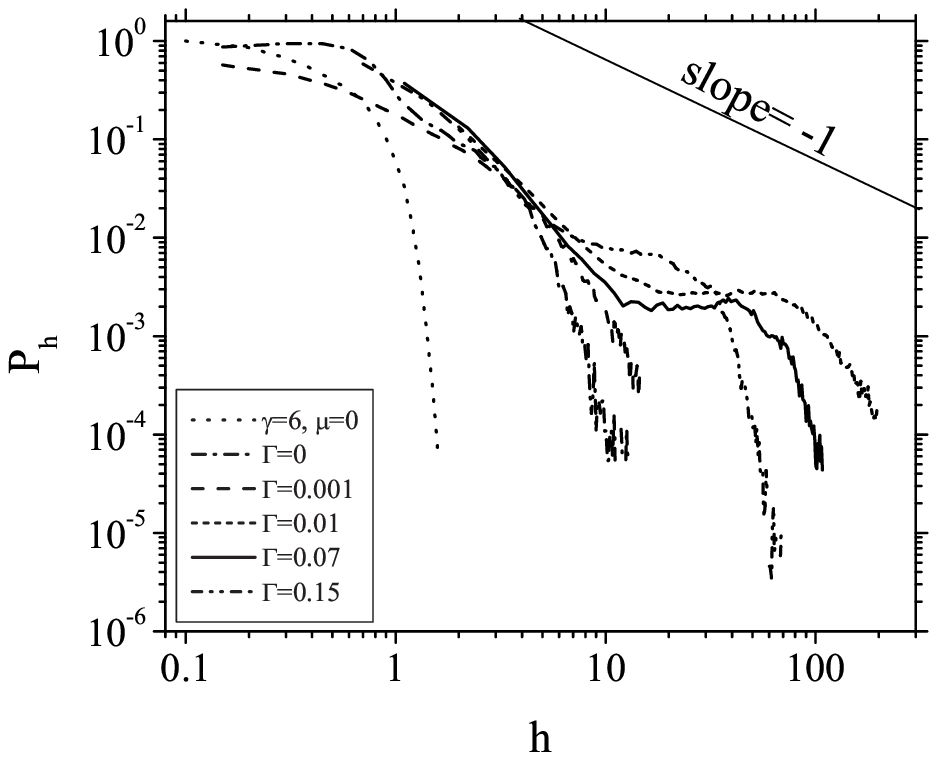,width=2.0in}\label{Fig. 2}}
%  \hspace*{-1.5cm}
  \minifigure[
The normalized probability for the appearance of extreme events
$P_{ee}$ as a function of the integrability parameter $\Gamma$. ]
     {\epsfig{figure=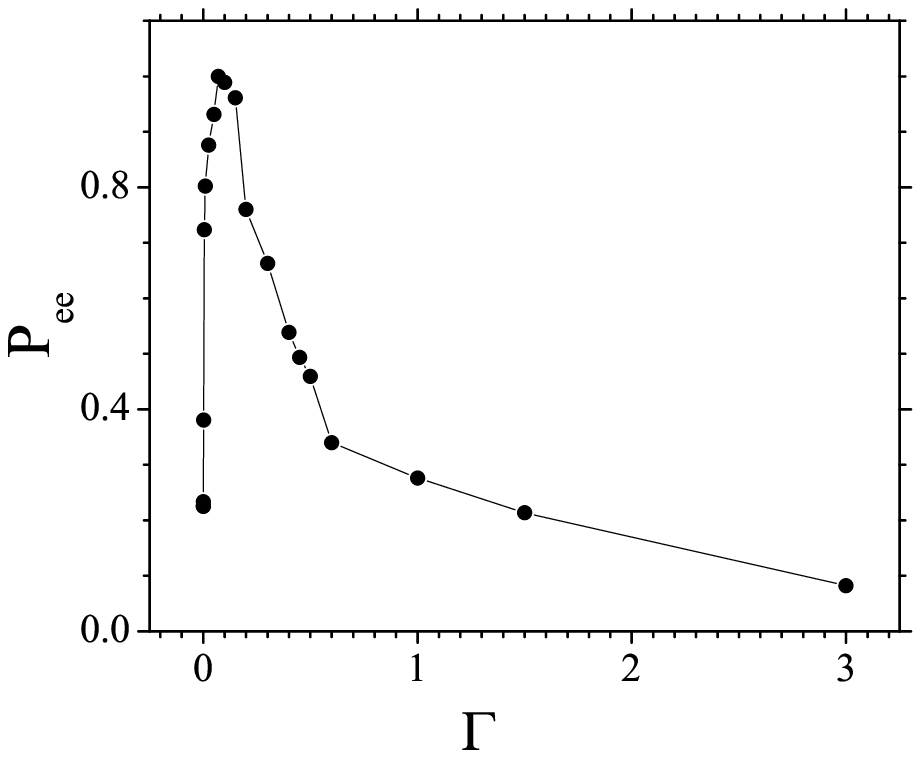,width=2.0in}\label{Fig. 3}}
}
\end{figure}
%%%%%%%%%%%%%%%%%%end-figure2-3%%%%%%%%%%%%%%%%%%%%%%%%%%%%%%%%%%%%%%%%%%%%%%%%%

The calculated height probability distributions (HPD) $P_h (h)$
shown in Fig. \ref{Fig. 2} are in accordance with the earlier
observations. The forward (backward) height $h$ at the $n-$th site
is defined as the difference between two successive minimum
(maximum) and maximum (minimum) values of $|\xi_n (t) |$. Both the
forward and backward heights are then used for the calculation of
the local height distribution; the HPDs are eventually obtained by
spatial averaging of all local height distributions. The tails of
the HPDs shown in the figure for finite $\Gamma$ are rather long,
while in some cases they form a plateau, indicating that EEs with
height several times that of the mean of the distribution are
probable to appear. In the DNLS limit ($\Gamma \gg 1$), the HPDs
are very close to a Rayleigh distribution which tails decay
exponentially \cite{VanKampen1981}, indicating negligible
probability for the appearance of EEs (dotted curve in Fig.
\ref{Fig. 2}).

In order to calculate the probability of appearance of EEs, we
adopt the criterion employed frequently in the context of water
waves and define as an EE a wave which height is greater than
$h_{th}$, where $h_{th}=2.2 h_s$, with $h_s$ being the significant
wave height. The latter is defined as the average height of the
one-third higher amplitude waves in the height distribution. The
probability for the appearance of EEs $P_{ee} = P_h (h>h_{th})$ is
then obtained by integration of the corresponding (normalized) HPD
from $h=h_{th}$ up to infinity. Following this procedure, the
probability of EE appearance, $P_{ee}$, is calculated as a function of $\Gamma$
(Fig. \ref{Fig. 3}). As can be observed in this figure, $P_{ee}$ has a
finite value in the integrable AL case ($\Gamma =0$) around $0.2$.
Then, it increases with increasing $\Gamma$ and forms a
resonance-like peak with maximum at $\Gamma \simeq 0.07$. Further
increase of $\Gamma$ leads to a decrease of $P_{ee}$, that becomes
vanishingly small for $\Gamma >>1$. This behavior is compatible
with the DB picture outlined earlier, which indicates that the
weakly nonlinear, nearly integrable regime is favorable for the
appearance of EEs.

\section{The two-dimensional DNLS model}

We consider the dynamics in a two-dimensional tetragonal lattice with diagonal
disorder, or, equivalently, wave propagation in a two-dimensional array of
evanescently coupled optical nonlinear fibers with random index variation,
both described through the disordered DNLS equation, viz.
\begin{equation}
\label{1}
i\frac{d{\psi}_{n,m}}{dt}=  {\epsilon}_{n,m} {\psi}_{n,m}+J({\psi}_{n+1,m}+{\psi}_{n-1,m}+
{\psi}_{n,m+1}+{\psi}_{n,m-1})
 +\gamma \left\vert {\psi}_{n,m}\right\vert ^{2}{\psi}_{n,m},
\end{equation}
where $n,m=1,...,N$, $\psi_{n,m}$ is a probability (or wave)
amplitude at site $(n,m)$ , $J > 0$ is the inter-site coupling
constant accounting  for tunnelling between adjacent sites of the
lattice (corr. evanescent coupling), $\gamma$ is the nonlinearity
parameter that stems from strong electron-phonon coupling (corr.
Kerr nonlinearity), while ${\epsilon}_{n,m}$, is the local site
energy (related to the fiber refractive index), chosen randomly
from a uniform, zero-mean distribution in the interval $[-W/2,+W/2]$.
Equation (\ref{1}) serves as a paradigmatic model for a
wide class of physical problems where both disorder and nonlinearity are present.
For $\gamma \rightarrow 0$, Eq.(\ref{1}) reduces to the 2D Anderson model
while in the absence of disorder (${\epsilon}_{n,m}=0$), it reduces to the DNLS
equation in two dimensions that is generally non-integrable.
Eq. (\ref{1})  conserves the norm
\begin{equation}
\label{100}
   P_N=\sum_{n=1}^N \sum_{m=1}^M |{\psi}_{n,m}|^2 ,`
\end{equation}
and the Hamiltonian $\cal{H}$, corresponding to total probability
(corr. input power)  and the energy of the system, respectively.
In optics, the sign of the nonlinearity strength $\gamma$ determines the
focussing ($\gamma >0$) or defocusing ($\gamma <0$) properties of the nonlinear
medium.

Eqs. (\ref{1}), implemented with periodic boundary conditions, are integrated
using a $6-$th order Runge-Kutta solver with fixed time-stepping \cite{Maluckov2009},
for several values of $W$ and $\gamma$ ($J=1$), for a lattice with $N=41$.
Larger lattices (i.e., with $N=81$) give similar results.
The system is initialized with a uniform state that is slightly modulated by
periodic perturbations in order to facilitate the development of MI 
\cite{Kivshar1994,Wollert2009}.
The MI threshold for Eq. (\ref{1}) in the absence of disorder can be obtained
using the standard linear analysis, as in the previous section.
The MI induces nonlinearly localized modes that are, however, modified by the
presence of the quenched disorder. In the time scale of the numerical study,
the presence of disorder induces additional energy redistribution among the
lattice sites. Thus, it weakens the energy self-trapping which in two dimensions
would result in strongly pinned and highly localized breathers \cite{Molina1999}.
%%%%%%%%figure4%%%%%%%%%%%%%%%%%%%%%%%%%%%%%%%%%%%%%%%%%%%%%%%%%%%%%%%%%%%%%%%%%
\begin{figure}[h!]
\center
\includegraphics[angle=0, width=1.0\linewidth]{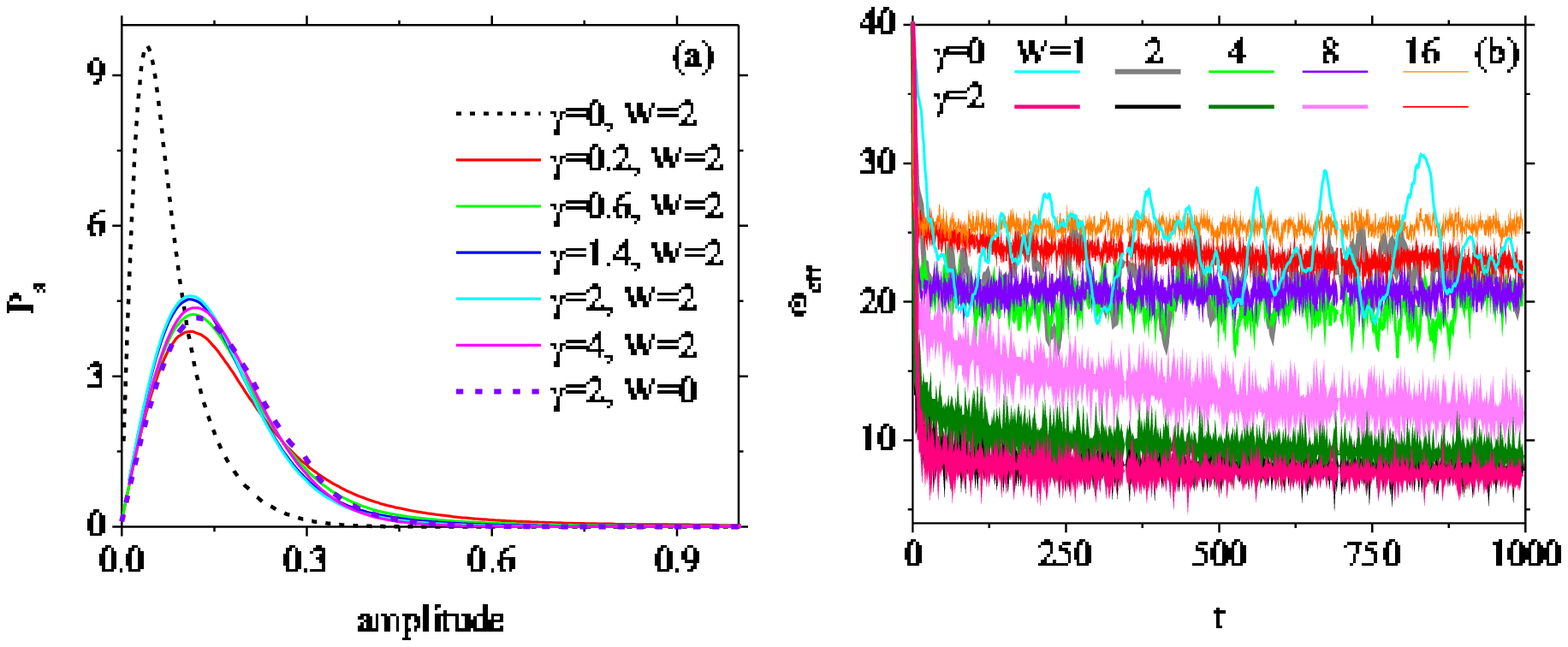}
\caption{(color online)
(a) Amplitude probability distributions as a function of amplitude $P_s (|\psi |)$
for several nonlinearity strengths and $W=2$.
(b) Effective localization length $w_{eff}$ of the localized structures formed
in the lattice as a function of time $t$ for several levels of disorder and
$\gamma =0,~2$.
} \label{fig1}
\end{figure}
%%%%%%%%end-figure4%%%%%%%%%%%%%%%%%%%%%%%%%%%%%%%%%%%%%%%%%%%%%%%%%%%%%%%%%%%%%

The criterion for defining an EE is the same as in the previous
section, with the obvious replacement of $|\xi_n (t)|$ by
$|\psi_{n,m} (t)|$ in the definition of the wave height $h$.
During relatively long time (typically $\sim 10^3$ time units or
equivalently approximately $500$ coupling lengths) the system
self-organizes and localized structures appear on different sites
that are surrounded by irregular, low-amplitude background. Some
of these structures are in the form of DBs, either pinned or
mobile, while some others are transient. The complete amplitude
statistics for the observed time interval is shown in Fig. \ref{fig1}a
for several levels of disorder $W$ and focusing
nonlinearity strengths $\gamma$; In all cases, we observe
Rayleigh-like distributions which parameters depend on $\gamma$
and $W$. Any state of the lattice that appears with probability in
the long tails of these distributions is a potential candidate for an EE.
In order to quantify the onset of EEs in the lattice,
several statistical measures have been used, viz. the probability
for the appearance of EEs, $P_{ee}$, the first appearance and
recurrence EE times, $R$ and $P_r$, respectively, as well as the
inverse participation ratio
\begin{equation}
\label{3}
   P={P_N^2} \left\{ \sum_{n=1}^{N}\sum_{m=1}^{M}|f_{n,m}|^4 \right\}^{-1} ,
\end{equation}
where $P_N$ is the norm. We may then define the effective localization length
\begin{equation}
\label{33}
   w_{eff}=P^{1/2} ,
\end{equation}
which provides the average spatial extent of the structures formed after
the development of MI (Fig. \ref{fig1}b).
Note that increasing strength of focusing nonlinearity ($\gamma >0$) decreases
$\omega_{eff}$ (enhances localization), while increasing the magnitude of the
defocusing nonlinearity increases $\omega_{eff}$ (reduces localization).
%%%%%%%%figure2%%%%%%%%%%%%%%%%%%%%%%%%%%%%%%%%%%%%%%%%%%%%%%%%%%%%%%%%%%%%%%%%%
\begin{figure}[h!]
\center
\includegraphics[angle=0, width=1.0\linewidth]{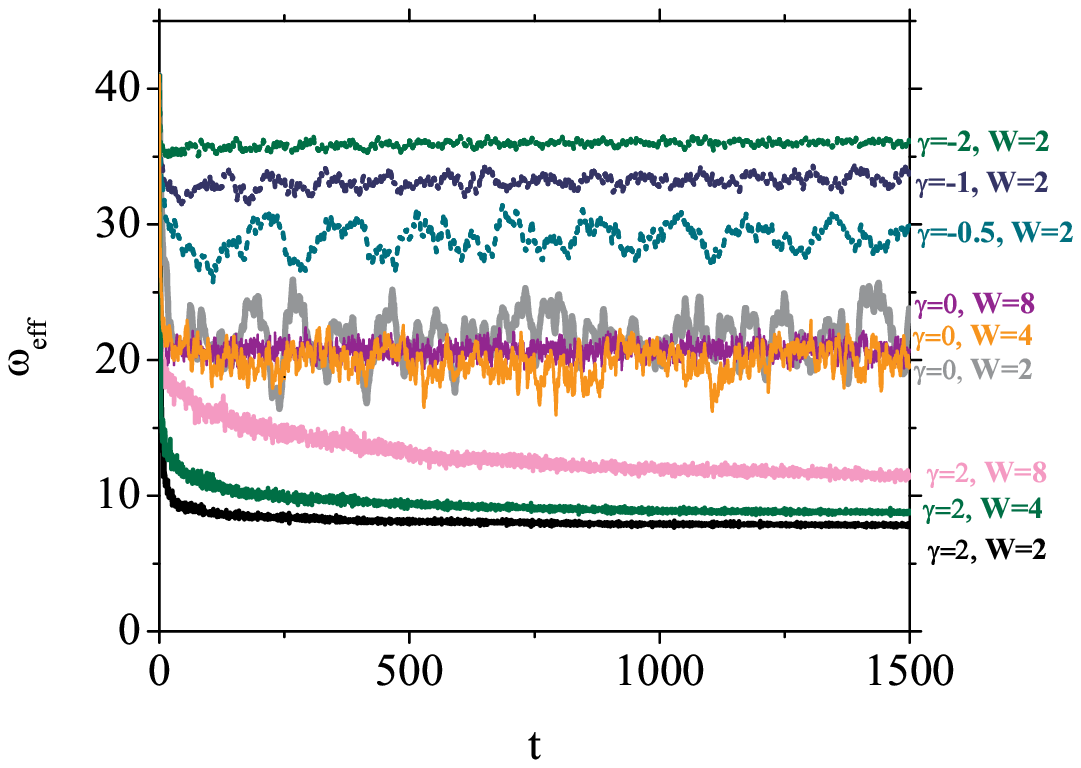}
\caption{(color online)
Comparison of the effective averaged localization length $\omega_{eff}$
as a function of time $t$ for several levels of disorder and both focusing and
defocusing nonlinearities.
} \label{fig2}
\end{figure}
%%%%%%%%figure2%%%%%%%%%%%%%%%%%%%%%%%%%%%%%%%%%%%%%%%%%%%%%%%%%%%%%%%%%%%%%%%%%

Both disorder and nonlinearity, each acting alone, favor wave localization in
the lattice. When they are simultaneously present, quenched disorder dominates
the early stage dynamics since MI develops slowly, at least for relatively small
nonlinearity strengths. In this regime, Anderson-like localized states decay
spatially while still permitting local energy redistribution until a lower-energy
stable localized state is reached. As it can be observed in Fig. \ref{fig1}b,
in the presence of nonlinearity, the effective localization length $w_{eff}$
saturates to a value lower than that for the corresponding linear lattice for a
wide range of disorder levels.
This tendency is compatible with the findings in Ref. \cite{Schwartz2007},
where it was also observed that increasing self-focusing strength enhances
localization. On the other hand, $w_{eff}$ increases with increasing level of
disorder, favoring de-localization.
The tendency of disorder-induced-delocalization in the presence of nonlinearity
can be attributed to the partial destruction of pinned, highly localized DBs
for relatively high levels of disorder.
In the case of defocusing nonlinearity (Fig \ref{fig2}), $w_{eff}$ increases
with increasing magnitude of the nonlinearity strength.
%%%%%%%%figure3%%%%%%%%%%%%%%%%%%%%%%%%%%%%%%%%%%%%%%%%%%%%%%%%%%%%%%%%%%%%%%%%%
\begin{figure}[h!]
\center
\includegraphics[angle=0, width=1.0\linewidth]{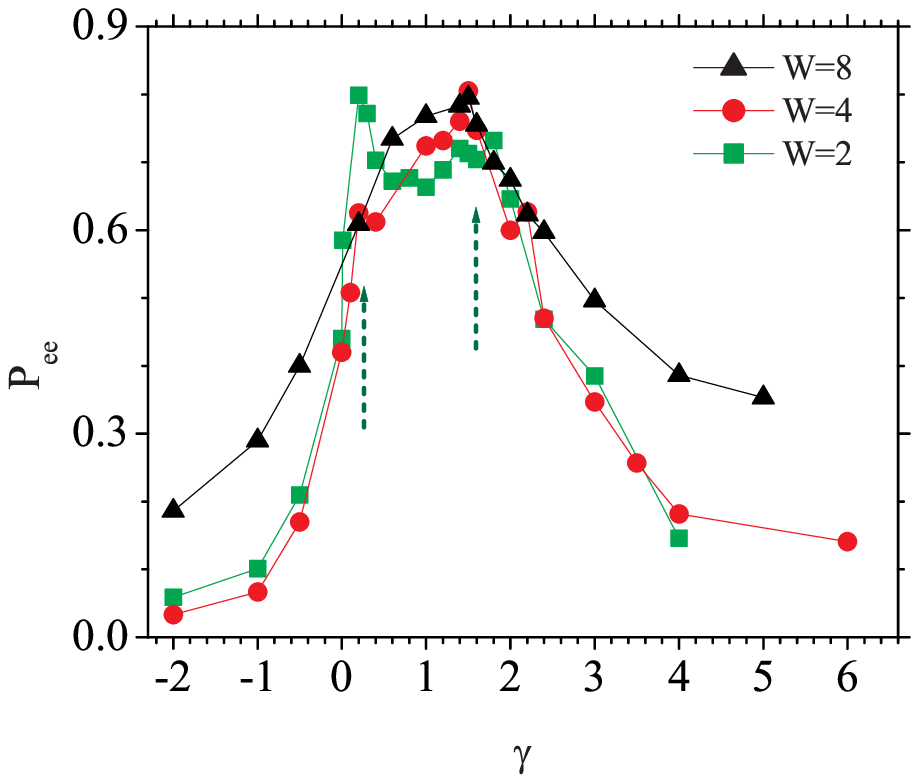}
\caption{(color online)
Extreme event height probability $P_{ee}$ as a function of the nonlinearity
strength $\gamma$ for several levels of disorder $W$.
} \label{fig3}
\end{figure}
%%%%%%%%figure3%%%%%%%%%%%%%%%%%%%%%%%%%%%%%%%%%%%%%%%%%%%%%%%%%%%%%%%%%%%%%%%%%

For obtaining the favorite parameter intervals for the appearance
of EEs, we calculate numerically $P_{ee}$ as in the previous
section as a function of $\gamma$ for three levels of disorder
(Fig. \ref{fig3}). Note that both negative and positive values of
$\gamma$ have been included in this figure. Referring to the case
of high disorder level, we observe that the probability $P_{ee}$
increases with $\gamma$ increasing from negative values, until it
reaches a maximum that is located at small positive $\gamma$.
Further increase of $\gamma$ decreases $P_{ee}$. For lower levels
of disorder, the calculated $P_{ee} (\gamma)$ dependencies exhibit
secondary maxima. Also, the  decrease of $P_{ee} (\gamma)$, for
$\gamma$ moving away from its values at the maxima on either side,
is much faster compared to that for high levels of disorder. For
zero nonlinearity $P_{ee}$ has still appreciable values; we obtain
$P_{ee}=0.47,~0.44$ for disorder levels with $W=2,~4$,
respectively. Thus, according to Fig. \ref{fig3}, the appearance
of EEs is favored in the part of parameter space that corresponds
to weak nonlinearity strengths $\gamma$ and moderate levels of
disorder $W$. For lower levels of disorder ($W=2, 4$), the first
local maximum can be correlated with the high $P_{ee}$ for the
nearly integrable lattice discussed in the previous section
\cite{Maluckov2009}. On the other hand, for arbitrary level of
disorder, the self-trapping effect of nonlinearity seems to be
responsible for the appearance of the second local maximum at
$\gamma \approx 2$ with $P_{ee}\approx 0.7\% $.
%%%%%%%%%%%%%%%%%%%%%%%%%%%%%%%%%%%%%%%%%%%%%%%%%%%%%%%%%%%%%%%%%%%%%%%%%%%%%%%

\begin{figure}[h!]
\center
\includegraphics[width=11cm]{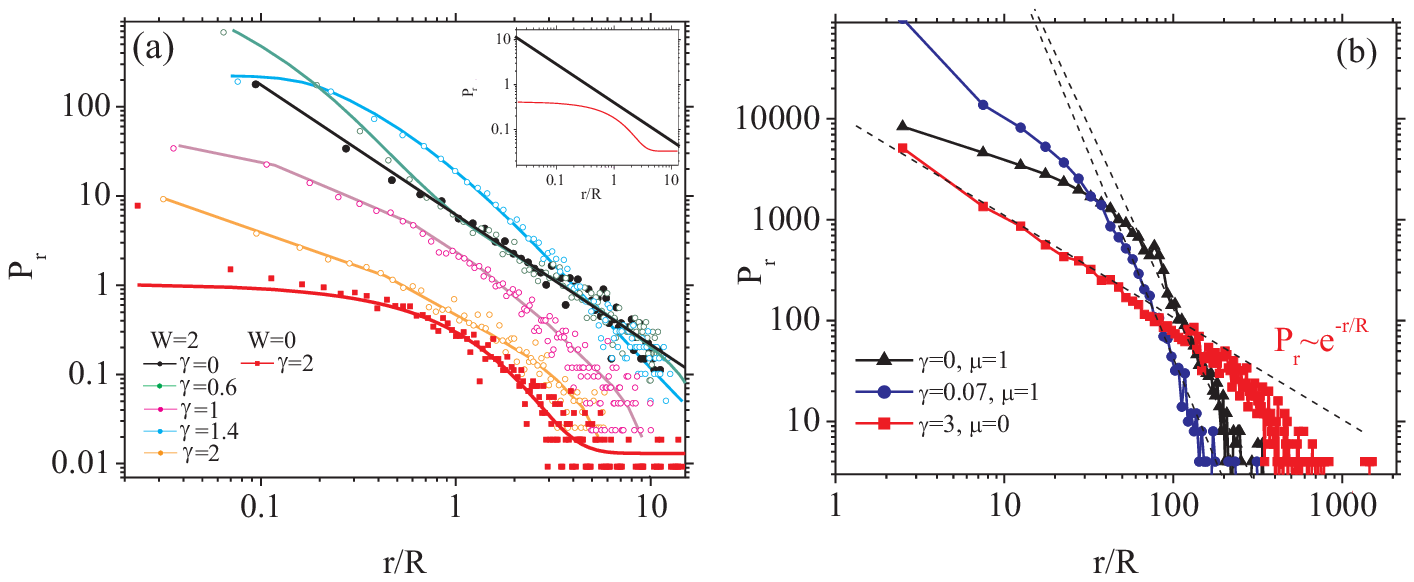}
\caption{(color online) (a) Return time probability $P_r$ (not
normalized) as a function of $r/R$ ($q=0.66$), for a linear
disordered lattice (power law fit - black curve), a nonlinear
ordered (exponential fit - red line) and intermediate cases
(double exponential fits). Inset: the first two cases are shown
separately. (b) The $P_r$ (not normalized) for the Salerno lattice
for parameters written in figure.} \label{fig5}
\end{figure}

For gaining deeper understanding of the appearance of EEs in 
disordered nonlinear lattices, we calculate the return time
probabilities $P_r$ as a function of the recurrence time $r$ of 
EEs, and the mean recurrence time $R$ of an EE as a function of
the wave height threshold $q$ \cite{Altmann2005,Santhanam2008},
for focusing nonlinearity. The slope of $R$ as a function of $q$
is smaller in a linear disordered lattice compared to that of a
nonlinear disordered lattice for any level of disorder \cite{Maluckov2013}. 
Also, $R$ increases with increasing $q$ for any $W$ and $\gamma$;
that increase is however faster for lower disorder level and stronger nonlinearity.
The regime of strong nonlinearity and low level of disorder favors the creation
of highly pinned, immobile DBs through self-trapping, increasing thus dramatically
the mean recurrence time $R$.
The return time probabilities $P_r$ are calculated for several values of $W$ and 
$\gamma$, and shown in Fig. \ref{fig5}. For a given threshold $q$ ($q>h_{th}$)
we scan the lattice to find an event at a given 
location with amplitude larger than $q$. We then register as
recurrence time $r$, the time interval between this event and a
subsequent one with amplitude larger than $q$ that appears at the
same location. We follow this procedure repeatedly up to maximum
time and construct distributions that are scaled by the average
return time $R \equiv R(q)$, like those shown in Fig. \ref{fig5}.
For linear lattices in the presence of disorder, $P_r$ as a
function of $r/R$ fits to a power-law function of the form
\begin{equation}
\label{333}
   P_r =\left[ a +b \left( \frac{r}{R} \right) \right]^{(-1/c)} ,
\end{equation}
with $1/c$ being  ~$1.34$ and ~$2.46$ for $q=0.66$ and $q=0.33$,
respectively. Generally speaking, the presence of nonlinearity
reduces the probability of EE appearance. Remarkably, the $P_r$
vs. $r/R$ curves in Fig. \ref{fig5} (a) make a transition from a
power-law, in the linear disordered regime, to a double
exponential for intermediate nonlinearity, to a single exponential
in the nonlinear ordered regime. This transition is linked to the
behavior of the tail of the corresponding amplitude probability
distributions. In addition we found that the exponential decay of
the curves is faster for higher nonlinearity strength. Note that
these observations are in accordance to the return time
distribution transition from power-law to exponentially like one
with increasing nonlinearity parameter in the 1D lattice model
presented in previous section (see Fig. \ref{fig5} (b)).

%%%%%%%%figure5%%%%%%%%%%%%%%%%%%%%%%%%%%%%%%%%%%%%%%%%%%%%%%%%%%%%%%%%%%%%%%%%%
%%%%%%%%figure5%%%%%%%%%%%%%%%%%%%%%%%%%%%%%%%%%%%%%%%%%%%%%%%%%%%%%%%%%%%%%%%%%
Recent work on the disordered DNLS equation has proposed a "phase diagram" that
points the different regimes of wave-packet spreading \cite{Laptyeva2010,Flach2010}.
Our relatively short-time results could be related to expected long-time
wave-packet spreading regimes summarized in these references.
Different regimes are obtained for

(i) $\delta > 2$, onset of self-trapping,

(ii) $d<\delta <2$, strong chaos, and

\begin{figure}[h!]
\centerline{
  \minifigure[(color online)
Different regimes of wave-packet spreading in the effective parameter
space $(W,d)$.
Lines represent regime boundaries $\delta\approx d$ and $\delta =2$.
Empty squares denote parameters for which the $P_{ee}$ maxima are
found. Lines with arrows show the direction of increasing $P_{ee}$.]
     {\epsfig{figure=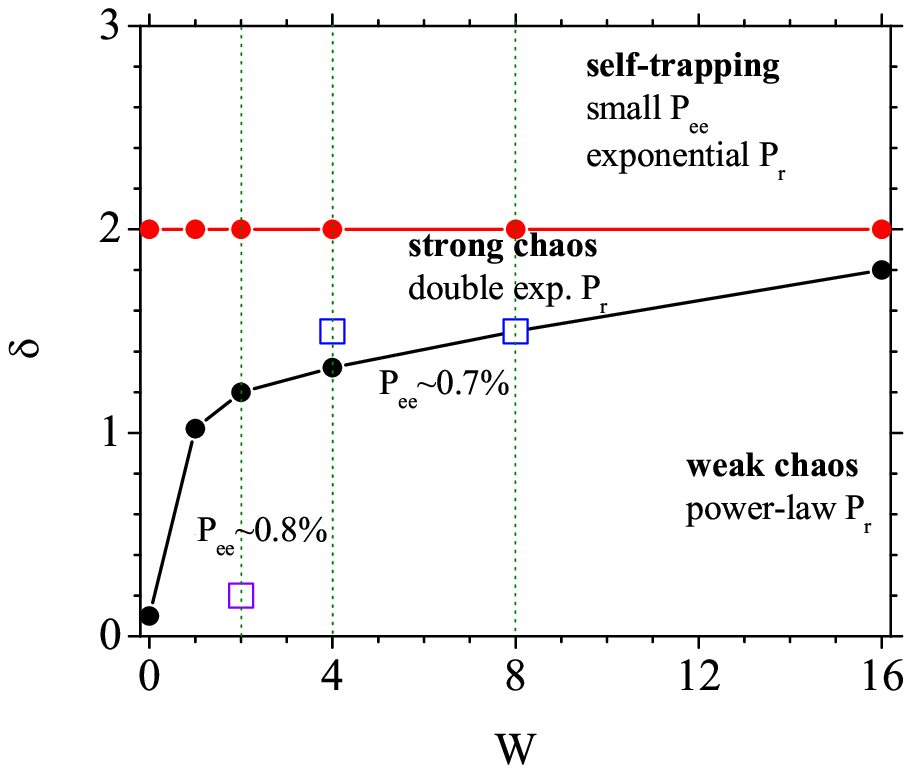,width=2.0in}\label{ra_fig6a}}
%  \hspace*{-1.5cm}
  \minifigure[Spatiotemporal distribution of the lattice energy:
disorder and self-focusing nonlinearity are $W=2,\,\gamma=2$ at
time $t=1900$.]
     {\epsfig{figure=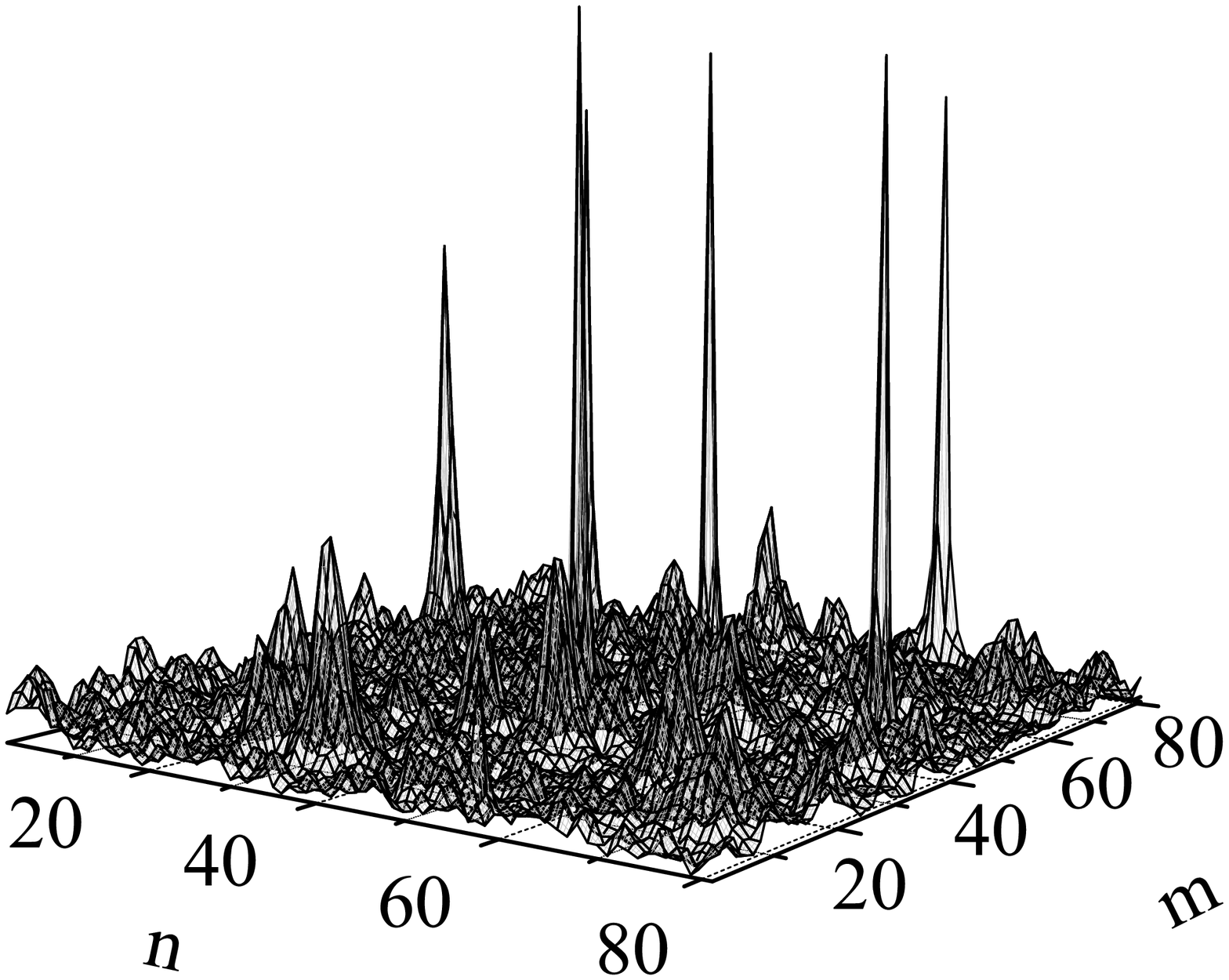,width=3.in}\label{ra_fig6b}}
}
\end{figure}

(iii) $\delta < d$,

\noindent where
\begin{equation}
\label{3333}
   d \approx \frac{\Delta}{V} =\frac{(8J+W)}{w_{eff}} , \qquad
   \delta \approx \gamma ,
\end{equation}
are the average frequency spacing of the nonlinear modes $\delta$
within a localization volume $V$, and the nonlinear frequency
shift, respectively. The selection of the regimes (i)-(iii) was
done taking into account the intensity of interaction among the
nonlinear modes; the latter increases with the nonlinearity
strength up to the high nonlinearity (here $\delta\approx 2$) when
the strong self-trapping results in the creation of isolated,
strongly pinned, high amplitude DBs. The comparison is only
approximate but leads to interesting observations (see Fig.
\ref{ra_fig6a}). The first local maximum in the $P_{ee}$ for weak
nonlinearity is located in the weak chaos regime. Its maximal
value $P_{ee}\approx 0.8\% $ is observed for small $W$. The
second, broader maximum of $P_{ee}$ for all levels of disorder is
located in the strong chaos regime relatively close to the border
lines with neighboring regimes. On the other hand, we may
associate the power-law decay of $P_r$ to the weak chaos regime,
and the exponential decay to the self-trapping regime. Therefore,
transient EEs are more probable in the regime of weak chaos, while
the long-lived EEs (high amplitude, strongly pinned DBs) dominate
for strong chaos and self-trapping. This enables us to relate the
first local maximum in $P_{ee}$ to the weak interaction of
transient EEs induced by disorder while the second, broader
maximum, to the appearance of longer-lived DBs resulting from the
energy redistribution through the strong interaction between
nonlinear modes. In Fig. \ref{ra_fig6b} we show a typical
spatiotemporal pattern generated in a nonlinear disordered
lattice. Note the presence several EEs, from which we can
distinguish at least six with very high amplitude.
%%%%%%%%figure9%%%%%%%%%%%%%%%%%%%%%%%%%%%%%%%%%%%%%%%%%%%%%%%%%%%%%%%%%%%%%%%%%
%%%%%%%%end-figure9%%%%%%%%%%%%%%%%%%%%%%%%%%%%%%%%%%%%%%%%%%%%%%%%%%%%%%%%%%%%%

\section{Conclusions.-}
Extreme events or rogue waves nowadays appear in many different
physical contexts, and their statistics deviates significantly
from the Gaussian behavior that was expected for random waves.
Among the many works that have appear the last few years
investigating different aspects about EE appearance and the
responsible mechanisms, relatively few of them are devoted to
discrete systems
\cite{Maluckov2009,Ankiewicz2010,Akhmediev2011,Maluckov2013}. In
particular, the role of integrability on the probability $P_{ee}$
of appearance of EEs and the wave amplitude distributions have
been explored in a model that interpolates between a
non-integrable and a completely integrable one. The power-law
dependence of the distributions reveals that the probability of EE
appearance is much more significant than expected from Gaussian
statistics. Moreover, the normalized $P_{ee}$ exhibits a
resonance-like maximum in the near-integrable limit. These results
can be further analyzed with the help of a nonlinear map
\cite{Maluckov2009}, where the onset of interaction between DBs
manifests itself as a transition from local to global
stochasticity monitored through the positive Lyapunov exponent.

In the presence of both disorder and nonlinearity there are two
processes that now act simultaneously; Anderson localization
\cite{Anderson1958} and self-focusing due to the MI of the CW
background.  When each of these processes proceeds alone, it
favors the formation of localized structures that eventually get
pinned in the lattice. When they act simultaneously, however, the
system exhibits higher complexity. In the first stages of the
evolution, disorder dominates the dynamics; at later stages,
however, nonlinearity takes over through the development of MI
that activates self-trapping mechanisms and tends to generate
pinned, high-amplitude localized structures which inhibit energy
exchange between lattice sites. For weak nonlinearity, however,
the pinning mechanism is not very effective, facilitating energy
exchange between sites and the appearance of high-amplitude,
localized, short-lived structures (EEs) at random locations. For
moderate levels of disorder, the probability of EE appearance is
maximized, giving the resonance-like peaks shown in Fig.
\ref{fig3}. The two peaks that are observed for weak nonlinearity,
are related to the weak and strong chaos regime. According to
previous analysis, the first local maximum in $P_{ee}$ can be
related to transient EEs (weak chaos regime), while the second one
to the formation of long-lived DBs. The passage from strong to
weak chaos, as well as from non-integrability to integrability, is
also related to the observed transition in $P_r$ vs. $r/R$ from an
exponential to a power-law.

%%%%% Acknowledgments %%%%%%%%%%%%%%%%%%%%%%%%%%%%%%%%%%%%%%%%%%%%%%%%%%%%%%%%%%
\section{Acknowledgments}
This research was partially supported by the European Union's Seventh Framework
Programme (FP7-REGPOT-2012-2013-1) under grant agreement n$^o$ 316165,
and by the Thales Project MACOMSYS,
cofinanced by the European Union (European Social Fund - ESF)
and Greek National Funds through the Operational Program
"Education and Lifelong Learning" of the National Strategic
Reference Framework (NSRF)  Research Funding Program: THALES.
"Investing in knowledge society through the European Social Fund".
A. M. and Lj. H. acknowledge support from the Ministry of Education,
Science and Technical Development of Republic of Serbia (Project III 45010).
%%%%% END-Acknowledgments %%%%%%%%%%%%%%%%%%%%%%%%%%%%%%%%%%%%%%%%%%%%%%%%%%%%%%

%%%%% References %%%%%%%%%%%%%%%%%%%%%%%%%%%%%%%%%%%%%%%%%%%%%%%%%%%%%%%%%%%%%%%
%\section{Bibliography}
%%%   \bibliographystyle{ws-rv-van}
%%%   \bibliography{../../Tex/My-BibTex-Library-05Oct2013}

%%%%% END-References %%%%%%%%%%%%%%%%%%%%%%%%%%%%%%%%%%%%%%%%%%%%%%%%%%%%%%%%%%%
%%%%%%%%%%%%%%%%%%%%%%%%%%%%%%%%%%%%%%%%%%%%%%%%%%%%%%%%%%%%%%%%%%%%%%%%%%%%%%%%
\end{document}